\documentclass[
reprint,
superscriptaddress,
showpacs,preprintnumbers,
amsmath,amssymb,
aip
]{revtex4-1}

\usepackage{color}
\usepackage{graphicx}
\usepackage{bm}
\usepackage{txfonts}
\usepackage{tikz}
\usepackage{pgffor}
\usepackage{verbatim}
\usepackage{pstricks}
\usepackage[colorlinks, breaklinks=true, linkcolor=blue, citecolor=blue, linktocpage=true]{hyperref}
\usepackage{epstopdf}

\begin{document}

\title{
Voltage-control of damping constant in magnetic--insulator/topological--insulator bilayers 
}
\date{\today}
\author{Takahiro Chiba}
 \affiliation{National Institute of Technology, Fukushima College, 30 Nagao, Kamiarakawa, Taira, Iwaki, 
Fukushima 970-8034, Japan}
\author{Alejandro O. Leon}
 \affiliation{Departamento de F\'isica, Facultad de Ciencias  Universidad Tecnol\'ogica Metropolitana, Las Palmeras 3360, \~Nu\~noa 780-0003, Santiago, Chile}
\author{Takashi Komine}
 \affiliation{Graduate School of Science and Engineering, Ibaraki University, 4-12-1 Nakanarusawa, Hitachi, Ibaraki 316-8511, Japan}
 
\begin{abstract}
The magnetic damping constant is a critical parameter for magnetization dynamics and the efficiency of memory devices and magnon transport. Therefore, its manipulation by electric fields is crucial in spintronics. Here, we theoretically demonstrate the voltage-control of magnetic damping in ferro- and ferrimagnetic--insulator (FI)/topological--insulator (TI) bilayers. Assuming a capacitor-like setup, we formulate an effective dissipation torque induced by spin-charge pumping at the FI/TI interface as a function of an applied voltage. By using realistic material parameters, we find that the effective damping for a FI with 10~nm thickness can be tuned by one order of magnitude under the voltage with 0.25~V. Also, we provide perspectives on the voltage-induced modulation of the magnon spin transport on proximity-coupled FIs.
\end{abstract}


\maketitle


Voltage or electric-field control of magnetic properties
is fundamentally and technologically crucial for energetically efficient spintronic technologies,~\cite{Ohno00,Song17} such as magnetic random-access memories (MRAMs),~\cite{Nozaki19} spin transistors,~\cite{Zutic04,Takiguchi19} and spin-wave-based logic gates.~\cite{Rana19}
In these technologies, voltage-control of magnetic anisotropy (VCMA) in thin ferromagnets~\cite{Weisheit07, Duan08, Maruyama09} promises energy-efficient reversal of magnetization by a pulsed voltage~\cite{Shiota12,Leon18,Chiba20} and manipulation of propagating spin waves with lower power consumption.~\cite{Rana19a} 
The control of magnetic damping is also highly desirable to increase the performance of spintronic devices. For instance, low magnetic damping allows small critical current densities for magnetization switching and spin-wave excitation by current-induced spin-transfer~\cite{STTReview} and spin-orbit torques.~\cite{Manchon19,Hamadeh14} On the other hand, a high magnetic damping can be beneficial in reducing the data writing time in MRAM devices. For magnonic devices, magnetic damping is a key factor because it governs the lifetime of spin waves or magnons as information carriers.~\cite{Chumak15}
Even if the magnetic damping is a vital material parameter that governs magnetization dynamics in several spintronic devices, its voltage-control is not fully explored except for a few experiments with ferro- and ferrimagnets.~\cite{Okada14,LChen15,Rana20,Xu19,Harder19} 

The main origin of magnetic dissipation is the spin-orbit interaction (SOI), which creates relaxation paths of the spin-angular momentum into conduction electrons and the lattice. Hence, potential candidates to achieve the voltage-control of magnetic damping are magnetic materials and/or strong SOI systems. Three-dimensional topological insulators (3D TIs), such as ${\rm Bi_2Se_3}$, are characterized by band inversion due to a strong SOI~\cite{Hasan10,Qi11} and possess an ideally insulating bulk and spin--momentum locked metallic surface states. Recently, ${\rm Bi_{2-x}Sb_{x}Te_{3-y}Se_{y}}$ (BSTS)~\cite{Ando13} and Sn-doped ${\rm Bi_{2-x}Sb_{x}Te_{2}S}$~\cite{Kushwaha16} have been reported to be ideal 3D TIs with two-dimensional (2D) Dirac electrons on the surface and a highly insulating bulk. For spintronics, the interface between a ferromagnet and a TI can enhance the magnitude of both spin and charge currents.~\cite{Mellnik14,Shiomi14} Some experiments reported~\cite{Jiang16,Wang16,Tang18,Fanchiang18} the spin-charge conversion at room temperature~\cite{Kajiwara10,Ding20} in a bilayer of TI/ferro- and ferrimagnetic-insulator (FI) such as Y$_3$Fe$_5$O$_{12}$~(YIG) with very low Gilbert damping constant ($\alpha$). An essential feature of the FI/TI bilayer is that the TI bulk behaves as a semiconductor, enabling the control of the surface carrier density by a voltage.~\cite{HWang19} Also, magnetically doped TI exhibit VCMA.~\cite{Fan16,Chiba20} Hence, TIs are a promising candidate to achieve the voltage-control of magnetic damping.

In this work, we theoretically demonstrate the voltage-control of magnetic damping in FI/TI bilayers. We formulate an effective dissipation torque induced by spin-charge pumping at the FI/TI interface as a function of a gate voltage $V_{\rm G}$. Our main result is that the voltage changes the effective damping by one order of magnitude for a FI with a perpendicular magnetization configuration and 10~nm thickness. Also, we provide perspectives on the modification of magnon scattring time in a FI-based magnonic device.

\begin{figure}[ptb]
\begin{centering}
\includegraphics[width=0.4\textwidth,angle=0]{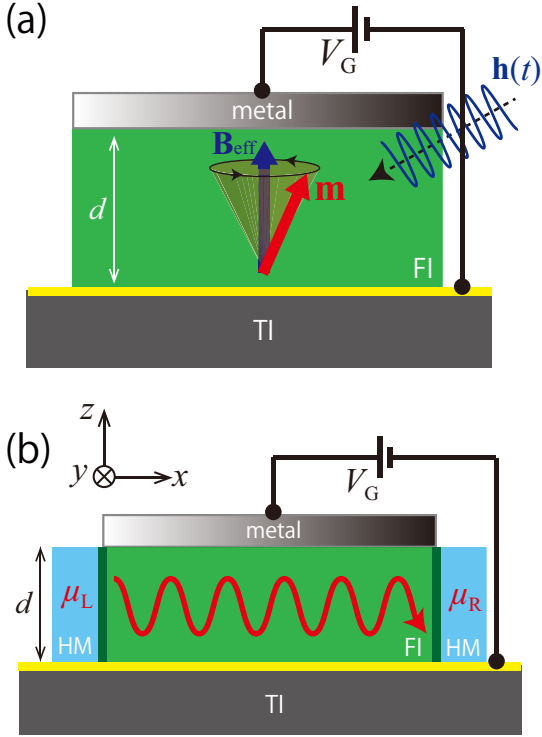} 
\par\end{centering}
\caption{
(a) Schematic geometry (side view) of a capacitor-like device comprising a ferromagnetic insulator (FI) film (with thickness $d$) sandwiched by a TI and a normal metal as a top electric gate with $V_{\rm G}$. The yellow line corresponds to the TI surface state. The red arrow denotes the precessional magnetization directions in the FMR driven by static (${\bf B}_{\rm eff}$) and oscillating (${\bf h}(t)$) magnetic fields.
(b) Schematic geometry of a transistor-like device comprising a FI film sandwiched by a TI and a normal metal. The red wave arrow represents a magnon current driven by the difference in spin accumulation ($\mu_{\rm L} - \mu_{\rm R}$) in the attached left and right heavy metal (HM) leads.
}
\label{fig:device}
\end{figure}
 

To study the effective damping torque, we consider 2D massless Dirac electrons on the TI surface with the magnetic proximity effect,~\cite{Jiang16,Wang16,Tang18,Fanchiang18} i.e., coupled to the magnetization of an adjacent FI. The exchange interaction between the surface electrons and the FI magnetization is modeled by a constant spin splitting along the magnetization direction with unit vector ${\bf m} = {\bf M}/M_{\rm s}$ (in which ${\bf M}$ is the magnetization vector with the saturation magnetization $M_{\rm s}$).~\cite{Nomura10} Then, the following 2D Dirac Hamiltonian provides a simple model for the FI/TI interface state~\cite{Chiba17}:
\begin{align}
\hat{H} = v_{\rm F}{\bm \sigma}\cdot\left(\hat{\bf p}\times\hat{\bf z}\right)+\Delta{\bm \sigma}\cdot{\bf m}, \label{H2d}
\end{align}
where $\hbar$ is the reduced Planck constant, $v_{\rm F}$ is the Fermi velocity of the Dirac electrons at zero applied voltage, $\hat{\bf p} = -i\hbar{\bm \nabla}$ is the momentum operator, $\{\mathbf{\hat{x}},\mathbf{\hat{y}},\mathbf{\hat{z}}\}$ are the unit vectors along the respective Cartesian axes, ${\bm \sigma} = (\sigma_x,\sigma_y,\sigma_z)$ is the vector of Pauli matrices for the spin, and $\Delta$ is the exchange interaction constant. For simplicity, we ignore here the particle--hole asymmetry and the hexagonal warping effect in the surface bands. 
Also, $\Delta$ and $v_{\rm F}$ are assumed to be temperature independent.~\cite{VFTIndep2} Note that we operate in the weak magnetic coupling limit, and therefore self-consistent treatment for the induced gap ($\Delta$)~\cite{Efimkin14} is not necessary.

Let us begin by calculating the dissipation torque induced by the spin-charge pumping~\cite{Yokoyama10,Sakai14,Shiomi14,Ndiaye17} of a dynamic magnetization in FI/TI bilayers. 
A precessing magnetization, driven by ferromagnetic resonance (FMR), as shown in Fig.~\ref{fig:device},
can be regarded as an effective vector potential ${\bf A}_{\rm eff}(t) = \Delta/(ev_{\rm F})\hat{\bf z}\times{\bf m}(t)$ with the electron charge $-e~(e>0)$, which drives a charge current via an effective electric field ${\bf E}_{\rm eff} =  -\partial_t {\bf A}_{\rm eff}$ (see the supplementary material), i.e.,
\begin{align}
{\bf J}_{\rm P} = \frac{\Delta}{ev_{\rm F}}\left(  \sigma_{\rm AH}\frac{\partial  \bf m}{\partial  t} - \sigma_{\rm L}\hat{\bf z}\times\frac{\partial  \bf m}{\partial  t}\right), \label{JP}
\end{align}
where $\sigma_{\rm L}$ and $\sigma_{\rm AH}$ are longitudinal and transverse (anomalous-Hall) conductivities, respectively, and depend on the $z-$component of the magnetization ($m_z$)~\cite{Chiba17}.
From the Hamiltonian~(\ref{H2d}), the velocity operator $\hat{\bf v} = \partial\hat{H}/\partial\hat{\bf p} = v_{\rm F}\hat{\bf z}\times{\bm \sigma}$ depends linearly on ${\bm \sigma}$. Therefore, the nonequilibrium spin polarization ${\bm \mu}_{\rm P}$ (in units of m$^{-2}$) is a linear function of the charge current ${\bf J}_{\rm P}$ on the TI surface, i.e., $
{\bm \mu}_{\rm P} = \hat{\bf z}\times{\bf J}_{\rm P}/(ev_{\rm F})$.
This nonequilibrium spin polarization ${\bm \mu}_{\rm P}$ exerts a dissipation torque on the magnetization, ${\bf T}_{\rm SP} = -\gamma\Delta/(M_{\rm s}d){\bm \mu}_{\rm P}\times{\bf m}$, namely
\begin{align}
{\bf T}_{\rm SP} &= \left(  -\alpha_{\rm AH}m_z + \alpha_{\rm L}{\bf m}\times\right)\left(  \frac{\partial \bf m}{\partial  t} - \frac{\partial  m_z}{\partial  t}\hat{\bf z}\right),
\label{EDT}
\end{align}
with
\begin{align}
\alpha_{\rm L(AH)} = \frac{\gamma\Delta^2}{e^2v_{\rm F}^2M_{\rm s}d}\sigma_{\rm L(AH)}, \label{alpha}
\end{align}
where $\gamma$ is the gyromagnetic ratio and $d$ is the thickness of the FI layer.
Equation~(\ref{EDT}) is  equivalent to the charge-pumping-induced damping-like torque that {\it Ndiaye}~{\it et~al.}\ derived using the Onsager reciprocity relation for a current-induced spin-orbit torque.~\cite{Ndiaye17} The first term in Eq.~(\ref{EDT}) originates from the magnetoelectric coupling (the Chern--Simons term)~\cite{Nomura10, Garate10} and renormalizes the gyromagnetic ratio. By using parameters listed in TABLE~\ref{tab.parameter}, $\Delta = 40$~meV,  and $d = 10$~nm, $\alpha_{\rm AH} \approx 10^{-4}$ is estimated even by using $\sigma_{\rm AH}$ at 0~K as the upper value.~\cite{Chiba17} Thus, we disregard the renormalization of the gyromagnetic ratio. In contrast, the second term in Eq.~(\ref{EDT}) stems from the Rashba--Edelstein effect due to the spin-momentum locking on the TI surface~\cite{Yokoyama10} and contributes to magnetic damping. 
Since we are interested in voltage-control of magnetic damping, we hereafter focus on $\alpha_{\rm L}$ in this study. 

\begin{figure*}[ptb]
\begin{centering}
\includegraphics[width=1.0\textwidth,angle=0]{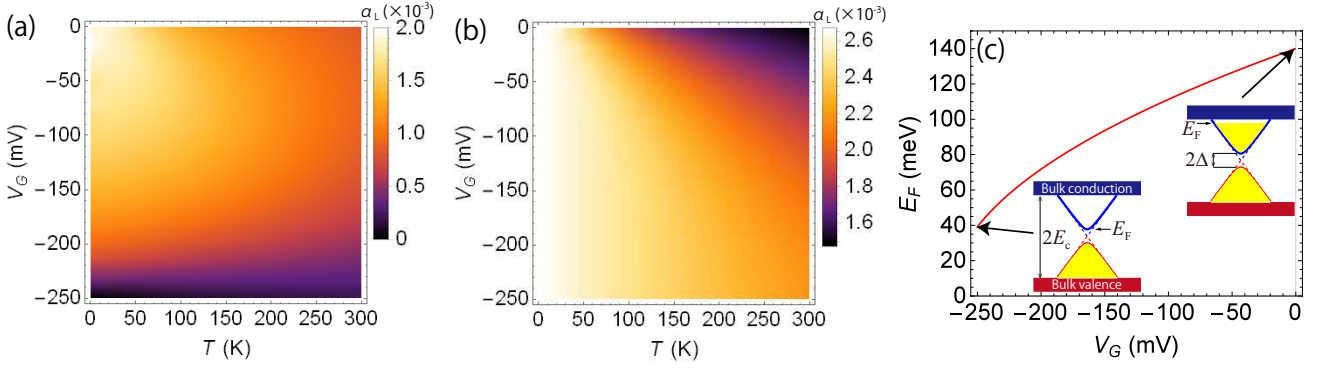} 
\par\end{centering}
\caption{
Effective damping enhancement $\alpha_{\rm L}$ of a TI/FI bilayer as functions of $V_{\rm G}$ and $T$ for $E_{\rm F}(V_{\rm G} = 0) = 140$~meV: (a) $m_z = 1$, (b) $m_z = 0$. (c) Voltage modulation of $E_{\rm F}$ in a TI. Insets represent schematic of massless (dashed line) and massive (solid line) surface state dispersions in the bulk band gap. In these graphs, we use parameters listed in Table~\ref{tab.parameter}, $\Delta = 40$~meV, and $d = 10$~nm for a FI thickness. The details of the calculations are given in the text. 
}
\label{fig:damping}
\end{figure*}

According to Eq.~(\ref{alpha}), the electric field effect on the conductivity $\sigma_{\rm L}$ can be used to control the magnetic dissipation. Namely, the voltage--induced change of the interfacial density of states in $\sigma_{\rm L}$ renders the TI a more or less efficient spin sink. The damping enhancement $\alpha_{\rm L}$ depends on the chemical potential $\mu$, measured from the original band-touching (Dirac) point. At room temperature, or below it,  the thermal energy is much smaller than the Fermi one, $k_{\rm B}T\ll E_{\rm F}$, with $T$ the temperature and $k_{\rm B}$ the Boltzmann constant. Then, we can use the following Sommerfeld expansion of the chemical potential $\mu$
\begin{align}
\mu(T)\approx E_{\rm F}\!\left[  1-\frac{\pi^2}{6}\left(  \frac{k_{\rm B}T}{E_{\rm F}}\right)^2 \right]\label{mu}
\end{align}
with the voltage-dependent Fermi energy,~\cite{Chiba20} $E_{\rm F}=\mu(0)$, given by
\begin{align}
E_{\rm F}(V_{\rm G}) = \hbar v_{\rm F}\sqrt{4\pi\left( n_{\rm int} + \frac{\Delta^2}{4\pi\left( \hbar v_{\rm F}\right)^2} + \frac{\epsilon}{ed}V_{\rm G}\right)}, \label{EVG} 
\end{align}%
where $\epsilon$ is the permittivity of a FI and $n_{\rm int}=\left(  E_{\rm F}^2(0)-\Delta^2\right)/\left(  4\pi\hbar^2v_{\rm F}^2\right)$ is the intrinsic carrier density, i.e., at $V_{\rm G} = 0$. Note that we can define a voltage-dependent surface electron density $n_V(V_G)\equiv n_\text{int}+\epsilon V_G/\left(ed\right)$ that shows the underlying mechanism behind the voltage-control of interfacial phenomena in insulating bilayers with surface carriers, which goes beyond topological materials. Namely, a voltage increases or decreases the effective electron density and therefore enhances or weakens all effects that depend on this density, including isotropic~\cite{RKKYVolt} and anisotropic\cite{DMIVolt} exchange interactions, emergence of magnetization in metals,~\cite{PlatVolt} perpendicular magnetic anisotropy,~\cite{Nozaki19,Maruyama09} and spin-orbit torques.~\cite{Chiba20} The voltage-generated change in the surface density is equivalent to an interfacial Fermi energy shift. In this work, we predict that the spin-charge pumping efficiency is also modulated, an effect that may also appear in usual FI$\vert$\textit{normal metal} bilayers since the spin-mixing conductance depends on the electronic density.~\cite{SpinPumpingCahaya}

We investigate the effect of electric-gate on the effective damping $\alpha_{\rm L}$ so that we assume hereafter that the low-energy Dirac Hamiltonian~(\ref{H2d}) is an accurate description for a momentum cut $k_{\rm c} = \sqrt{E_{\rm c}^2 - \Delta^2}/(\hbar v_{\rm F})$, in which $2E_{\rm c}$ is the bulk bandgap of TIs~\cite{Tserkovnyak15}  (see Fig.~\ref{fig:damping}~(c)). Sufficiently far from the Dirac point ($\hbar\tau/E_{\rm F} \ll 1$, $\tau$ is the transport relaxation time), the electron scattering can be treated by the first Born approximation.~\cite{Adam09} With this, the longitudinal conductivity reads~\cite{Chiba19}
\begin{align}
\sigma_{\rm L} 
 = \frac{e^2}{2h} \int_{-E_{\rm c}}^{E_{\rm c}}dE_k \frac{E_k\tau(E_k,T)}{\hbar} \frac{E_k^2 - \Delta^2 m_{z}^2}{E_k^2 + 3\Delta^2 m_{z}^2}\left(  -\frac{\partial f_{\rm FD}}{\partial E_k}\right)
,\label{sigma L}
\end{align}
where $f_{\rm FD} = \left[  \exp{\left\{  (E_k - \mu)/(k_{\rm B}T)\right\}} + 1\right]^{-1}$ is the Fermi-Dirac distribution, the energy $E_k$ is the eigenvalue of Eq.~(\ref{H2d}), and $\tau(E_k,T)$ is the transport relaxation time of massless Dirac electrons within the Born approximation for impurity and phonon scatterings. By applying the Matthiessen rule,
\begin{align}
\frac{1}{\tau(E_k,T)} = E_k\left(  a + bk_{\rm B}T\right)
,\label{tau}
\end{align}
where $a = nV_0^2/\left( 4\hbar^3 v_{\rm F}^2\right)$~(in units of ${\rm eV^{-1}s^{-1}}$) parameterize contribution of the impurity scattering,~\cite{Ivanov18,Giraud12} $n$ is the impurity concentration, and $V_0$ is the scattering potential. Also, contribution to the transport relaxation time from the phonon scattering~\cite{Ivanov18,Giraud12} can be approximated by $b = D_0^2/\left(  4\hbar^3 v_{\rm F}^2\rho  t_{\rm s} v_{\rm L}^2\right)$~(in units of ${\rm eV^{-2}s^{-1}}$), where $\rho$ is the mass density of the quintuple layer (QL) in the TI crystal structure, $t_{\rm s}$ is the thickness of one atomic layer in 1~QL of TIs, $v_{\rm L}$ is the longitudinal phonon velocity, and $D_0$ is the deformation potential constant.

\begin{table}[ptb]
\begin{ruledtabular}\caption{\label{tab.parameter}Material parameters for the TI/FI bilayer.}
\begin{tabular}{ccccc}
& Symbol & Value & Unit \\ \hline
\footnotemark[1]BSTS Fermi velocity & $v_{\rm F}$ & $4.0\times10^5$ & ${\rm ms^{-1}}$ \\
\footnotemark[1]BSTS bulk band gap & $2 E_{\rm c}$ & 300 & meV \\ \hline
\footnotemark[2]YIG gyromagnetic ratio & $\gamma$ & 1.76$\times 10^{11}$ & ${\rm T^{-1}s^{-1}}$ \\
\footnotemark[2]YIG Gilbert damping constant & $\alpha$ & 6.7$\times 10^{-5}$ &  \\
\footnotemark[2]YIG saturation magnetization & $M_{\rm s}$ & 1.56$\times 10^5$ & ${\rm Am^{-1}}$ \\
\footnotemark[3]YIG relative permittivity & $\epsilon/\epsilon_0$ & 15 & 
\end{tabular}
\end{ruledtabular}
\footnotemark[1]{Reference~\onlinecite{Ando13},}
\footnotemark[2]{Reference~\onlinecite{Kajiwara10},}
\footnotemark[3]{Reference~\onlinecite{Sadhana09}.}
\end{table}

\begin{figure}[ptb]
\begin{centering}
\includegraphics[width=0.4\textwidth,angle=0]{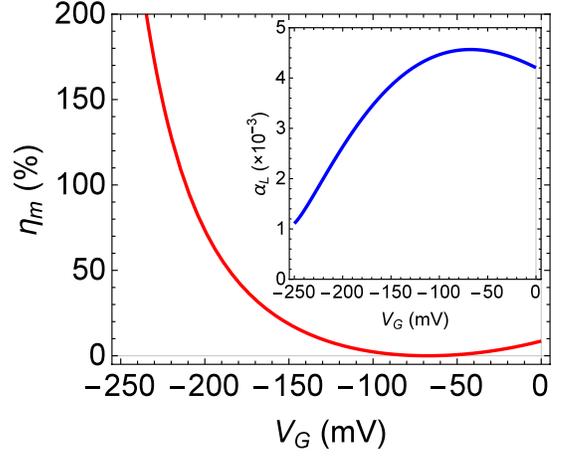} 
\par\end{centering}
\caption{
Modulation efficiency ($\eta_{\rm m}$) as a function of the gate voltage. Inset shows the corresponding behavior of the effective damping enhancement $\alpha_{\rm L}$. In these graphs, we use parameters listed in Table~\ref{tab.parameter}, $\Delta = 40$~meV, and $d = 2$~nm for a FI thickness.  We also set $E_{\rm F}(V_{\rm G} = 0) = 140$~meV and $T = 300$~K. 
}
\label{fig:efficiency}
\end{figure}

Figures~\ref{fig:damping}~(a) and (b) show the $V_{\rm G}$ and $T$ dependence of the effective damping enhancement $\alpha_{\rm L}$ for out-of-plane ($m_z = 1$) and in-plane ($m_z = 0$) magnetization configurations, respectively. 
Also, Fig.~\ref{fig:damping}~(c) illustrates the voltage modulation of $E_{\rm F}$ in TI. The bulk damping constant can be influenced by material and device parameters, such as SOI and magnetic anisotropies~\cite{Ding20}. However,  we predict the voltage-modulation of the \textit{damping enhancement} by spin-charge pumping. Therefore, our results are independent of the intrinsic dissipation mechanisms. At the FI/TI interface, orbital hybridization between TI and the 3$d$ transition metal in FI, such as YIG, deforms the TI surface states, which might shift the Dirac point to the lower energy and lift up $E_{\rm F}$,~\cite{Marmolejo17} so that we consider relatively high value $E_{\rm F}(V_{\rm G} = 0) = 140$~meV with the corresponding carrier density of the order of $10^{12}$~cm$^{-2}$. Also, $\Delta$ is used within the values reported experimentally in FI-attached TIs.~\cite{Hirahara17, Mogi19} For impurity parameters, we use $n = 10^{11}$~cm$^{-2}$ and $V_0 = 0.15$~keV\AA$^2$ based on an analysis of the transport properties of a TI surface.~\cite{Chiba19} We could not find estimates of the phonon scattering for BSTS in the literature so that we adopt those of non-substituted Bi$_2$Te$_3$ being $v_{\rm L} = 2.9\times10^5$~${\rm ms^{-1}}$, $D_0 = 35$~eV, $t_{\rm s} = 0.16$~nm, and $\rho = 7.86\times10^{3}$~${\rm kgm^{-3}}$ in Ref.~\onlinecite{Huang08}. These scattering parameters describe a relatively clean interface with the sheet resistance $\sim1$~k$\Omega$, which is one order less than that of experiments. In Figs.~\ref{fig:damping}~(a) and (b), $\alpha_{\rm L}$ monotonically decreases with increasing $T$ at a fixed $V_{\rm G}$ while it has peaks for changing $V_{\rm G}$ at a fixed $T>0$ (see also the inset of Fig.~\ref{fig:efficiency}). This feature reflects thermal excitation of surface carriers into the bulk states ($E_k > E_{\rm c}$), reducing the spin-charge-pumping contribution. With the out-of-plane configuration, $\alpha_{\rm L}$ can be tuned by one order of magnitude under the voltage, while $\alpha_{\rm L}$ changes by less than a factor two with the in-plane state, which suggests that the out-of-plane configuration is superior in controllability. 
The calculated $T$ dependence of damping enhancement at $V_{\rm G} = 0$ for the in-plane configuration agrees with a few experiments with the FI/TI bilayer.~\cite{Tang18,Liu20} 
Note that at much lower than $E_{\rm F}(V_{\rm G} = -250~{\rm mV}) \approx 40$~meV, our calculation with the in-plane configuration breaks down because of the finite level broadening due to the higher-order impurity scattering.~\cite{Shon98} The $V_{\rm G}$--dependent FMR is characterized by the Landau-Lifshitz-Gilbert theory in the supplementary material.


The electric manipulation of magnon spin transport is a relevant topic in spintronics. For example, in YIG with an injector and a detector Pt contact, 
changes of the magnon spin conductivity can be obtained by using a third electrode that changes the magnon density,~\cite{Cornelissens2018,EControlOfMagnTransp,ByAHEFF} potentially providing a functionality similar to the one a {\it field-effect transistors}. Damping compensation by current-driven torques~\cite{DampComp1,DampComp2} in magnetic heterostructures also influences magnon transport. Here, we provide a perspective on the electric-field-induced modulation of magnon scattering time, $\tau_{\rm m}$. 
Magnons can be injected and detected by their interconversion with charge currents in adjacent heavy metals (HMs) through the direct and inverse spin-Hall effects.~\cite{Kajiwara10} Similar to charge transport induced by an electrochemical potential gradient, a magnon spin current can be driven by the gradient of a magnon chemical potential injected by an external source.~\cite{Cornelissen16,Basso16} Magnon transport through a FI can be controlled by the gate voltage that modulates the effective damping in Eq.~(\ref{alpha}). 

So far, the magnon spin transport in the FI/TI bilayer lacks microscopic theory with few exceptions.\cite{Okuma17,Imai18} However, from the bulk of magnon spin transport,~\cite{Cornelissen16} the control of $\tau_{\rm m}$ results in the modification of all transport properties, including the magnon spin conductivity. In the presence of a TI contact, interfacial magnons are scattered by conducting Dirac electrons on the TI surface.~\cite{Yasuda16}  Considering a very thin ferromagnet that can be modeled by a 2D magnet. The inset of Fig.~\ref{fig:efficiency} shows that the damping enhancement is at least one order of magnitud larger than the bulk one of YIG.~\cite{Kajiwara10,Ding20} Accordingly, let us assume that interfacial magnons are absorbed by transferring their energy and angular momentum to Dirac electrons at a rate $1/\tau_{\rm m} \propto \alpha_{\rm L}$.~\cite{Cornelissen16} While there is no know microscopic expression for the magnon spin conductivity in the present system, bulk magnon transport obeys the relationship $\sigma_{\rm m} \propto\tau_{\rm m}$,~\cite{Cornelissens2018,EControlOfMagnTransp}  where $\sigma_{\rm m}$ is the magnon spin conductivity. In our case, the scattering time $\tau_{\rm m}$ is dominated by the magnon-relaxation process into the FI/TI interface. 
To estimate an effect of electric-gate on the magnon spin transport, we define the modulation efficiency
\begin{equation}
\eta_{\rm m} 
= \frac{\tau_{\rm m}(V_{\rm G}) - \tau_{\rm m}(V_{\rm max})}{\tau_{\rm m}(V_{\rm max})} 
= \frac{\alpha_{\rm L}(V_{\rm max})}{\alpha_{\rm L}(V_{\rm G})} - 1
,\label{efficiency}
\end{equation}
where $V_{\rm max}~(\approx -68$~mV for Fig.~\ref{fig:efficiency}) gives the maximum value of $\alpha_{\rm L}$ (and therefore the minimum value of $\tau_{\rm m}$).
In principle, $\tau_{\rm m}$ depends on $V_{\rm G}$ through not only $\alpha_{\rm L}$ but also via magnon dispersion relation, $\hbar\omega_{\bf q}$,~\cite{Cornelissen16} including a $V_{\rm G}$--dependent magnetic anisotropy. However, this $V_{\rm G}$--dependence is quite small even for a FI with 2~nm thickness (see the supplementary material), so that we disregard the influence of the magnon gap in the following calculation.
Figure~\ref{fig:efficiency} shows $V_{\rm G}$-dependence of the modulation efficiency at room temperature in which the strongly nonlinear behavior is interpreted as follows. Down to $V_{\rm G} \approx -130~{\rm mV}$, $\alpha_{\rm L}$ is affected by the thermal excitation of surface carriers, which makes a peak around $V_{\rm G} \approx -70~{\rm mV}$. From $-130~{\rm mV}$ to $-250~{\rm mV}$, the thermal excitation is suppressed, so that $\alpha_{\rm L}$ monotonically decreases with $|V_{\rm G}|$ due to the reduction of the Fermi surface. Hence, in this regime, one can effectively modulate the magnon spin transport by the voltage. 

In summary, we have theoretically demonstrated the voltage-control of magnetic damping in ferro- ferrimagnetic insulator (FI)/topological insulator (TI) bilayers. Assuming a capacitor-like setup, we formulate an effective damping torque induced by spin-charge pumping at the FI/TI interface as a gate voltage function. The presence of a perpendicular electric field results in a shift of the Fermi level or, equivalently, a modified interfacial electron density, increasing or decreasing the efficiency of the pumping process. We studied the consequences of this damping enhancement using realistic material parameters for FI and TI. We found that the effective damping with the out-of-plane magnetization configuration can be modulated by one order of magnitude under the voltage with 0.25~V. 
The present results motivate an application: the magnon scattering time can be tuned by a gate voltage, potentially allowing for a magnon transistor type of application. A complete quantitative description of the latter requires a microscopic theory of magnon spin transport in FI/TI bilayers, which might remain an unexplored issue.
The voltage-control of magnetic damping paves the way for low-power spintronic and magnonic technologies beyond the current-based control.

\

See the supplementary material for the calculation of the spin-charge pumping in FI/TI bilayers, the characterization of the FMR under several values of the applied voltage, the influence of $V_{\rm G}$--dependence of the anisotropy in the magnon dispersion.

\
We thank Camilo Ulloa and Nicolas Vidal-Silva for fruitful discussions.
This work was supported by Grants-in-Aid for Scientific Research (Grant No.~20K15163  and No.~20H02196) from the JSPS and Postdoctorado FONDECYT 2019 Folio 3190030.

\section*{Data Availability}

The data that support the findings of this study are available from the corresponding author upon reasonable request.





\end{document}